\documentclass[epj]{svjour}

\usepackage{graphicx}
\usepackage{fancyhdr}
\def\makeheadbox{{%  remove EPJC header
 }}
\def\mytitle{My title} 
\def\myauthors{My name}  
\def\mytype{My type of session}
\def\mysession{My session}
\def\mytitle{Branching Fractions and Direct CP Asymmetries
of Charmless B Decay Modes at CDF} %Put your title here!
\def\myauthors{Michal Kreps}    %Put your name here!
\def\mytype{Contributed Talk}    
\def\mysession{Flavor Physics}
\usepackage{lineno}

\newcommand{\ACPddef}{\ensuremath{{\frac{\BR (\aBdKpi)-\BR
(\BdKpi)}{\BR (\aBdKpi)+\BR (\BdKpi)}}}}
\newcommand{\ACPsdef}{\ensuremath{{\frac{\BR (\aBsKpi)-\BR
(\BsKpi)}{\BR (\aBsKpi)+\BR (\BsKpi)}}}}
\newcommand{\rateratiodef}{\ensuremath{\frac{\mathit{f_d}}{\mathit{f_s}}{\frac{ \Gamma(\aBdKpi)-
\Gamma(\BdKpi)}{\Gamma(\aBsKpi)-\Gamma(\BsKpi)}}}}

\newcommand{\BdpipisuBdKpidef}{\ensuremath{\frac{\BR(\Bdpipi)}{\BR(\BdKpi)}}}
\newcommand{\BsKKsuBdKpidef}{\ensuremath{\frac{\mathit{f_s}}{\mathit{f_d}}\frac{\BR(\BsKK)}{\BR(\BdKpi)}}}

\newcommand{\BsKpisuBdKpidef}{\ensuremath{\frac{\mathit{f_s}}{\mathit{f_d}}\frac{\BR(\BsKpi)}{\BR(\BdKpi)}}}
\newcommand{\BspipisuBdKpidef}{\ensuremath{\frac{\mathit{f_s}}{\mathit{f_d}}\frac{\BR(\Bspipi)}{\BR(\BdKpi)}}}
\newcommand{\BdKKsuBdKpidef}{\ensuremath{\frac{\BR(\BdKK)}{\BR(\BdKpi)}}}

\newcommand{\LbppisuLbpKdef}{\ensuremath{\frac{\BR(\Lbppi)}{\BR(\LbpK)}}}

\newcommand{\Bdpipi}{\ensuremath{\bd \to \pi^+ \pi^-}}

\newcommand{\BdKpi}{\ensuremath{\bd \to K^+ \pi^-}}
\newcommand{\aBdKpi}{\ensuremath{\abd \to K^- \pi^+}}
\newcommand{\BsKpi}{\ensuremath{\bs \to K^- \pi^+}}
\newcommand{\aBsKpi}{\ensuremath{\abs\to K^+ \pi^-}}
\newcommand{\BsKK}{\ensuremath{\bs \to  K^+ K^-}}

\newcommand{\Bspipi}{\ensuremath{\bs \to  \pi^+ \pi^-}}

\newcommand{\BdKK}{\ensuremath{\bd \to  K^+ K^-}}

\newcommand{\Lbppi}{\ensuremath{\Lambda_{b}^{0} \to p\pi^{-}}}

\newcommand{\LbpK}{\ensuremath{\Lambda_{b}^{0} \to pK^{-}}}

\newcommand{\bd}{\ensuremath{B^{0}}}        %   B-zero
\newcommand{\bs}{\ensuremath{B^{0}_s}}        %   B (zero) sub-s
\newcommand{\abd}{\ensuremath{\overline{B}^{0}}}    %   B-zero
\newcommand{\abs}{\ensuremath{\overline{B}^{0}_s}}    %   B (zero) sub-s
\newcommand{\BR}{\ensuremath{\mathcal B}}
\newcommand{\CP}{CP}            % CP

\newcommand{\acpbdkpi}{\ensuremath{A_{\rm{\CP}}(\bdkpi)}}

\newcommand{\acpbskpi}{\ensuremath{A_{\rm{\CP}}(\bskpi)}}

\newcommand{\bdkpi}{\ensuremath{\bd \to K^+ \pi^-}}

\newcommand{\bskpi}{\ensuremath{\bs \to K^- \pi^+}}

\newcommand{\Lb}{$\Lambda_{b}^{0}$}

\begin{document}
%%%%%%%%%%%%%%%%%%%%%%%%%%%%%%%%%%%%%%%%%%%%%%
% Toggle double spacing and line numbering
% Won't work with the PRD revtex4 !
%\pagewiselinenumbers
%\linenumbers
%%%%%%%%%%%%%%%%%%%%%%%%%%%%%%%%%%%%%%%%%%%%%%%

%
\title{Branching Fractions and Direct CP Asymmetries of
Charmless B Decay Modes at CDF}
\author{Michal Kreps\inst{1}\thanks{\emph{Email:}
kreps@ekp.uni-karlsruhe.de} on behalf of the CDF
Collaboration
%\thanks{\emph{Email:} kreps@ekp.uni-karlsruhe.de}%
}                     % Do not remove
\institute{Institute f\"ur Experimentelle Kernphysik,
Universi\"at Karlsruhe (TH), Postfach 6980, 76128 Karlsruhe,
Germany}
\date{}
\abstract{
We present new CDF results on the branching fractions and
time-integrated
direct CP asymmetries for $B^0$ and $B^0_s$ decay modes into
pairs of charmless
charged hadrons (pions or kaons). The data-set for this
update amounts to
$1$ $\mathrm{fb}^{-1}$ of $p\overline p$ collisions at a
center of mass energy $1.96$ $\mathrm{TeV}$. We report
the first observation of the $B_s \rightarrow K^- \pi^+$ mode and a
measurement of
its branching fraction and direct CP asymmetry. We also
observe for the
first time two charmless decays of the $\Lambda_b^0$-baryon:
$\Lambda_b^0 \rightarrow p\pi^-$ and $\Lambda_b^0 \rightarrow
pK^-$.
 % end of PACS codes
} %end of abstract
\maketitle
\section{Introduction}
\label{intro}

The decay modes of $B$ mesons into pairs of charmless
pseudo-scalar mesons are effective probes of the CKM matrix
with sensitivity to potential new physics. The production
cross section of $B$ hadrons at the Tevatron
allows studies of such decays which are competitive to B
factories. In addition $B$ hadrons of all
kind are produced at the Tevatron and thus making possible to complement the
information by studies of decays of $B_s$ mesons and $B$
baryons which is impossible at current B factories.

In this paper we review current results on charmless two
body decays of $B$ hadrons from the CDF experiment using $1$
$\mathrm{fb^{-1}}$ of data. 
We use charged kaons, pions and protons as as final state
particles.
The main results presented here are
the first observation of the decay $B_s^0\rightarrow K^-\pi^+$ and
measurement of direct CP asymmetries in $B^0\rightarrow
K^+\pi^-$ and $B_s^0\rightarrow K^-\pi^+$ decays. The
presented analysis is an update of the previous result
obtained with $360$ $\mathrm{pb^{-1}}$ of data
\cite{Abulencia:2006psa} and
more details about it can be found in the Ref. \cite{public_note}.

\section{Data sample}
We analyze a data sample with an integrated luminosity of $1$
$\mathrm{fb^{-1}}$ collected by the CDF II detector at
Tevatron. Online selection of events requires two
oppositely charged tracks, each with $p_T>2$
$\mathrm{GeV}/c$. The scalar sum $p_T(1)+p_T(2)$ is required to
be larger than $5.5$ $\mathrm{GeV}/c$ and the transverse opening
angle between tracks ranges from $20^\circ$ to $135^\circ$.
The CDF detector has the unique opportunity to use the silicon
tracking at trigger level. This allows us to place requirements on
the impact parameter $d_0$ of each track and on the
displacement of the secondary vertex $L_{xy}$. We use $100$ $\mathrm{\mu
m}<d_0<1$ $\mathrm{mm}$ for the two tracks, $d_0<140$
$\mathrm{\mu m}$ and $L_{xy}>200$ $\mathrm{\mu m}$ for the
$B$ hadron candidate.

The offline selection is done using pseudoexriments with
statistical uncertainty of the studied quantity as a figure
of merit.
During the optimization, the requirements from
the trigger level are tightened. In addition to tightening
trigger cuts we add other two discriminating variables,
which are the quality of the $B$ secondary candidate vertex fit and
the isolation defined as $I(B)\,=\,p_T(B)/[p_T(B)+\sum_i
p_T(i)]$, where the sum runs over all tracks not associated with
the $B$ candidate in a cone of unit
radius in the $\eta$--$\phi$ space around the $B$ candidate.
As a result we derive two sets of requirements, one
optimized for measurements in decay modes with larger
yields and the other one for measurements in modes with lower
yields. They optimize the sensitivity for the measurement of the CP
asymmetry $A_{\mathrm{CP}}(B^0\rightarrow K^+\pi^-)$ and for
the discovery of the decay $B_s^0\rightarrow K^-\pi^+$.

\section{Signal decomposition}

\begin{figure}[htb]
\centering
\includegraphics[width=0.43\textwidth,angle=0]{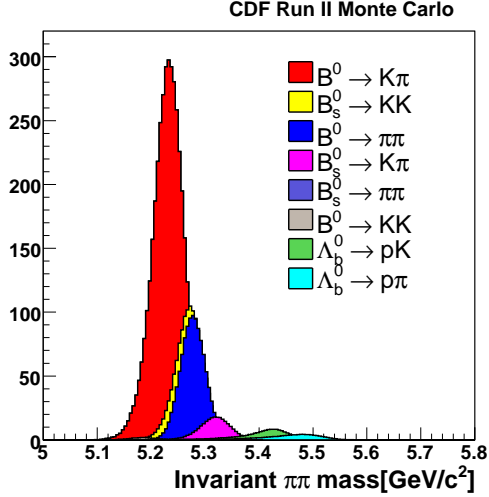}
\caption{Expected invariant mass distribution for different
charmless two-body decays of $B$ hadrons. In all cases the
invariant mass distribution is calculated assigning pion
mass to both decay products.}
\label{fig:1}
\end{figure}

In Fig.~\ref{fig:1} we show the typical expectation of the
invariant mass distribution obtained from simulated events
assigning pion mass to both decay products.
Despite the excellent mass resolution on the level of $22$
$\mathrm{MeV}/c^2$ it is not possible to distinguish
different decays as separate peaks in the invariant mass
distribution. In addition the particle identification system of
the CDF detector does not allow track-by-track
identification.   

\begin{figure}
\centering
\includegraphics[width=0.43\textwidth,angle=0]{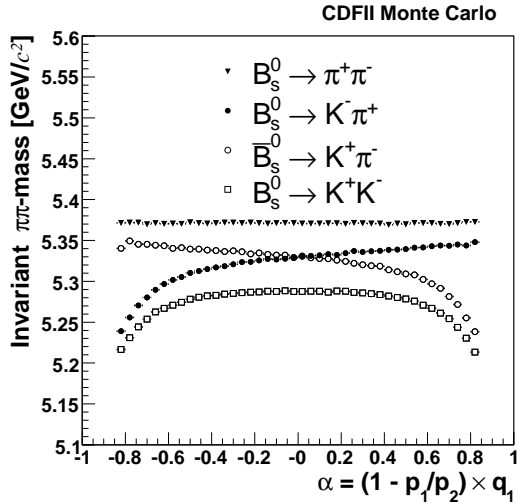}
\caption{Average $M_{\pi\pi}$ versus $\alpha$ for simulated
$B_s^0$ decays. Corresponding dependencies for other $B$ hadrons
are analogous.}
\label{fig:2}
\end{figure}

To find out the composition of the signal an unbinned maximum
likelihood fit is performed using the invariant $\pi\pi$ mass
$M_{\pi\pi}$,
$\mathrm{d}E/\mathrm{d}x$ measurement and momenta of the two tracks.
The fit determines fractions of different components, which
are then converted to physical quantities like branching
fractions and CP asymmetries.
The momenta of the two tracks enters in a form of loosely
correlated variables $p_{tot}=p_1+p_2$ and
$\alpha=(1-p_1/p_2)q_1$ where index 1 (2) corresponds to
particle with lower (higher) momentum.
The invariant mass with correct mass assignment to the
daughters can be unambiguously calculated from $M_{\pi\pi}$
and momenta of the two daughters.
The dependence of the $M_{\pi\pi}$ on the $\alpha$ for
$B_s^0$
decays is shown in Fig.~\ref{fig:2}. Decays of other three $B$
hadrons have similar dependencies.
The likelihood used in the fit has the form
 \begin{eqnarray}\label{eq:likelihood}
    \mathcal{L}_i & = & (1-b)\sum_{j} f_j
\mathcal{L}^{\mathrm{kin}}_j  \mathcal{L}^{\mathrm{PID}}_j
\nonumber \\
                  &   & +  b \left( f_{\rm{A}}
\mathcal{L}^{\mathrm{kin}}_{\mathrm{A}}
    \mathcal{L}^{\mathrm{PID}}_{\mathrm{A}}+
   (1-f_{\rm{A}}) \mathcal{L}^{\mathrm{kin}}_{\mathrm{E}}
    \mathcal{L}^{\mathrm{PID}}_{\mathrm{E}}
  \right) 
\end{eqnarray}
where $b$ is a background fraction and indices $A$ and $E$
label physics and combinatorial background. Index $j$ runs
over signal $B^0$, $B_s^0$ and $\Lambda_b^0$ decays modes. 
$\mathcal{L}^{kin}$ ($\mathcal{L}^{PID}$) denotes the mass
(particle identification) part of the likelihood.
The background mass distribution is described by an Argus
function convoluted with a Gaussian for physics background and
an exponential function for combinatorial background.

The need of disentangling of many overlapping signal
contributions demands a good description of the signal PDF
distributions in the likelihood fit.
One of the main effects for the mass description is the inclusion
of a tail due to the final state radiation. This is modeled using QED
calculations \cite{Baracchini:2005wp}. The result is then
convoluted with the resolution function obtained from
simulated events. The description is
found to be in excellent agreement with 
$D^0\rightarrow K^-\pi^+$ decays from
$D^{*+}\rightarrow D^0\pi^+$ decays reconstructed in the data.

For a proper description of the particle identification, we
calibrate the $\mathrm{d}E/\mathrm{d}x$ response using a data
sample of $D^{*+}\rightarrow D^0\pi^+$ with $D^0\rightarrow
K^-\pi^+$, which provides a very clean sample of kaons and
pions. On this sample we obtain a separation  between kaons and pions around
$1.4\sigma$. After the 
calibration we use the same data sample to derive the probability
density function for the likelihood. This is the same for signal and
background except of fractions of different particles, which
are independent for signal and background.

\section{Results}

\begin{figure}[htb]
\centering
\includegraphics[width=0.43\textwidth,angle=0]{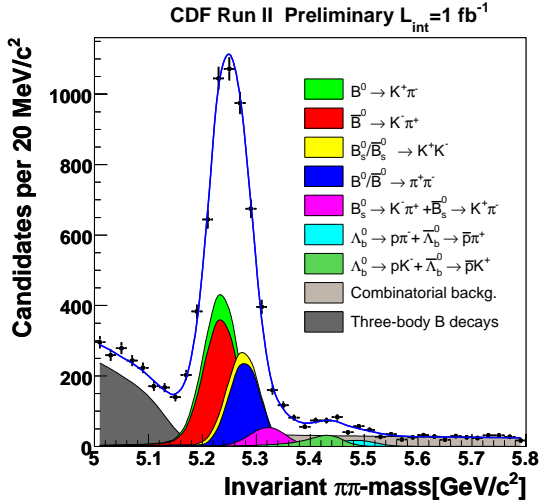}
\caption{Invariant mass distribution using selection
optimized for $B_s^0\rightarrow K^-\pi^+$ observation. The full
line represents result of the fit.}
\label{fig:3}
\end{figure}

The result of the analysis is a refined measurement of several
ratios of branching fractions for previously observed decays.
More important new decay modes $B_s^0\rightarrow K^-\pi^+$,
$\Lambda_b^0\rightarrow pK^-$ and $\Lambda_b^0\rightarrow p\pi^-$ 
are observed here for the first time. 
The invariant mass distribution using a set of
tighter requirements optimized for the observation of the
$B_s^0\rightarrow K^-\pi^+$ is shown in Fig.~\ref{fig:3}.
Measured ratios of the branching fractions for all observed
decay modes are summarized in Table~\ref{tab:br}. The
dominant systematic uncertainties are due to the statistical
uncertainty on selection efficiency, the uncertainty on
the $\mathrm{d}E/\mathrm{d}x$ calibration and parametrization
and the uncertainty on the background modeling.
In the following we will concentrate in more details on the two
results, which are the measurement of the ${B^0\rightarrow K^+\pi^-}$ CP
asymmetry and the observation of the $B_s^0\rightarrow
K^-\pi^+$ decay and measurement of the corresponding direct CP
asymmetry.
\begin{table*}
\caption{\label{tab:br}Results on the data sample optimized to measure \acpbdkpi\ (top) and
 \BR(\BsKpi) (bottom). Absolute branching fractions are normalized to the the world--average values
${\mathcal B}(\mbox{\BdKpi}) = (19.7\pm 0.6) \times 10^{-6}$ and
$f_{s}= (10.4 \pm 1.4)\%$ and $ f_{d}= (39.8 \pm 1.0)\%$~\cite{HFAG06}.
The first quoted uncertainty is statistical, the second is systematic.
$N_s$ is the number of fitted events for each mode. For rare
modes both the systematic and statistical uncertainty on $N_s$ was quoted
while for abundant modes only the statistical one. For the \Lb\ modes only the ratio \LbppisuLbpKdef\ was measured.}
{\footnotesize
\begin{tabular}{lc|lc|c}
\hline
Mode & N$_{s}$ & Quantity & Measurement & \BR (10$^{-6}$)  \\
\hline
\BdKpi         & 4045 $\pm$ 84          &  \ACPddef\         & -0.086 $\pm$ 0.023 $\pm$ 0.009   &                            \\
\Bdpipi        & 1121 $\pm$ 63          & \BdpipisuBdKpidef\ & 0.259 $\pm$ 0.017 $\pm$ 0.016    & 5.10 $\pm$ 0.33 $\pm$ 0.36 \\
\BsKK          & 1307 $\pm$ 64          & \BsKKsuBdKpidef\   &  0.324 $\pm$ 0.019 $\pm$ 0.041   & 24.4 $\pm$ 1.4 $\pm$ 4.6   \\
\hline
\BsKpi         & 230 $\pm$ 34 $\pm$ 16  & \BsKpisuBdKpidef\  &  0.066 $\pm$ 0.010 $\pm$ 0.010   & 5.0 $\pm$ 0.75 $\pm$ 1.0   \\
               &                        &  \ACPsdef\         &  0.39 $\pm$ 0.15 $\pm$ 0.08      &                            \\
               &                        &  \rateratiodef\    &  -3.21 $\pm$ 1.60 $\pm$ 0.39     &                            \\
\Bspipi        & 26 $\pm$ 16 $\pm$ 14   &\BspipisuBdKpidef\  &  0.007 $\pm$ 0.004 $\pm$ 0.005   & 0.53 $\pm$ 0.31 $\pm$ 0.40 \\
         &                        &              &                      & ($< 1.36$ @~90\%~CL)       \\
\BdKK          & 61 $\pm$ 25  $\pm$ 35  & \BdKKsuBdKpidef\   &  0.020 $\pm$ 0.008 $\pm$ 0.006   & 0.39 $\pm$ 0.16 $\pm$ 0.12 \\
         &                        &              &                            &  ($< 0.7$ @~90\%~CL)       \\
\LbpK          & 156 $\pm$ 20 $\pm$ 11  &\LbppisuLbpKdef\    &  0.66 $\pm$ 0.14 $\pm$ 0.08      &                            \\
\Lbppi         & 110 $\pm$ 18 $\pm$ 16  &                    &                                  &                            \\
\hline
\end{tabular}
}
\end{table*}

\subsection{$\mathbf{B^0\rightarrow K^+\pi^-}$ CP asymmetry}

\begin{figure}[hbt]
\centering
\includegraphics[width=0.43\textwidth,angle=0]{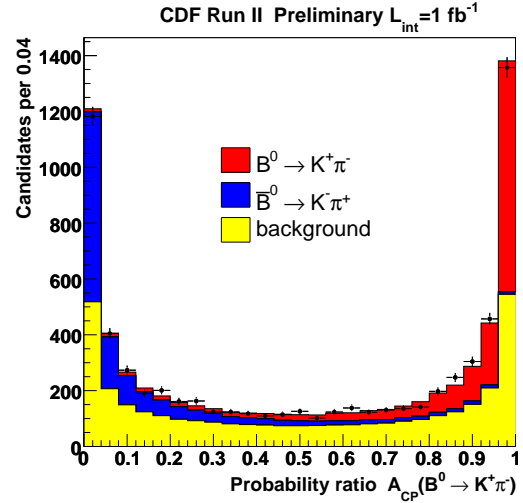}
\caption{Distribution of the probability ratio
$\mathcal{L}_{s1}/(\mathcal{L}_{s1}+\mathcal{L}_{s2})$ where
$\mathcal{L}_{s1}$ ($\mathcal{L}_{s2}$) denotes the probability
to be $B^0$ ($\overline{B}^0$). The points show data while
histograms represent different fit components.}
\label{fig:4}
\end{figure}

\begin{figure}
\centering
\includegraphics[width=0.40\textwidth,angle=0]{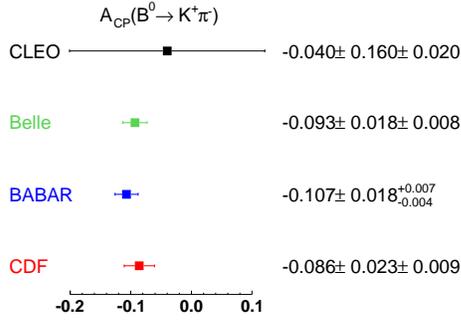}
\caption{Comparison of the measured $A_{\mathrm{CP}}(B^0\rightarrow
K^+\pi^-)$ with other measurements
\cite{Chen:2000hv,Belle:CPV,Aubert:2007mj}.}
\label{fig:5}
\end{figure}

As the decay $B^0\rightarrow K^+\pi^-$ is self tagging, one
can measure the direct CP asymmetry defined as
\begin{equation}
A_{\mathrm{CP}}\,=\,\frac{N(\overline{B}^0\rightarrow
K^-\pi^+) - N(B^0\rightarrow K^+\pi^-)}
{N(\overline{B}^0\rightarrow K^-\pi^+) + N(B^0\rightarrow
K^+\pi^-)}
\end{equation}
The only significant difference between $B^0$ and
$\overline{B}^0$ from the efficiency point of view is the difference
in the interaction of $K^+$ and $K^-$ with detector
material. This difference is estimated using a sample of
promt $D^0\rightarrow K^-\pi^+$ decays
and results in a $\le 0.6\%$ shift for the
$A_{\mathrm{CP}}$ obtained from the fit result.
To visualize the difference between $B^0$ and
$\overline{B}^0$ in Fig.~\ref{fig:4} we show the distribution of
the probability ratio
$\mathcal{L}_{s1}/(\mathcal{L}_{s1}+\mathcal{L}_{s2})$ where
$\mathcal{L}_{s1}$ ($\mathcal{L}_{s2}$) denotes the probability
to be $B^0$ ($\overline{B}^0$). The points show data while
histograms represent different fit components.
A small difference between $B^0$ and $\overline{B}^0$ is clearly visible.
The result corrected for detector effects is
$A_{\mathcal{CP}}\,=\,-0.086\pm0.023\pm0.009$. In
Fig.~\ref{fig:5} we show a comparison of this measurement to
the other existing measurements \cite{Chen:2000hv,Belle:CPV,Aubert:2007mj}. 
Our result is in good
agreement with other measurements with a precision comparable
to the Belle and {\it BABAR} experiments.

\subsection{$\mathbf{B_s^0\rightarrow K^-\pi^+}$ branching
fraction and CP asymmetry}

The most important result obtained here, is the first
observation of the decay $B_s^0\rightarrow K^-\pi^+$. We
observe $230\pm34\pm16$ signal events from which we measure 
\begin{equation} 
 \BsKpisuBdKpidef\ =\ 0.066\,\pm\,0.010\,\pm\,0.010.
\end{equation}
The significance of the observed signal is $8.2\sigma$
including systematic uncertainties. Using world average
values for $f_s$, $f_d$ and ${\mathcal B}(\mbox{\BdKpi})$
\cite{HFAG06} we
obtain for the branching fraction $\BR(\BsKpi)\,=\,(5.0\pm0.75\pm1.0)\cdot 10^{-6}$
which is in agreement with the latest theoretical
predictions \cite{Williamson:2006hb}.

\begin{figure}
\centering
\includegraphics[width=0.43\textwidth,angle=0]{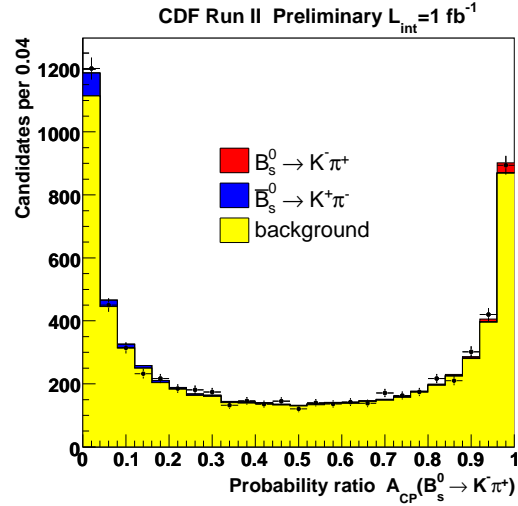}
\caption{Distribution of the probability ratio
$\mathcal{L}_{s1}/(\mathcal{L}_{s1}+\mathcal{L}_{s2})$ where
$\mathcal{L}_{s1}$ ($\mathcal{L}_{s2}$) denotes the probability
to be $B_s^0$ ($\overline{B}_s^0$). The points show data while
histograms represent different fit components.}
\label{fig:6}
\end{figure}

As the decay \BsKpi{} is a self-tagging decay, we can 
determine also the direct CP asymmetry. 
%This gives unique
%opportunity to CDF for a model independent test for new physics
%by comparing \acpbdkpi{} with \acpbskpi. 
The CDF experiment has the unique opportunity for a model
independent test for new physics by comparing \acpbdkpi{}
with \acpbskpi. In the standard model
one expects for the decay rate differences
\cite{Gronau:2000md} %\cite{Lipkin:2005pb}
\begin{equation}
\frac{\Gamma(\overline{B}^0\rightarrow
K^-\pi^+)-\Gamma(B^0\rightarrow K^+\pi^-)}{
\Gamma(B_s^0\rightarrow
K^-\pi^+)-\Gamma(\overline{B}_s^0\rightarrow K^+\pi^-)}=1.
\end{equation}
This can be used to predict \acpbskpi{} from the
known \acpbdkpi, ratios of branching fractions and
lifetimes. Using world average values provided by HFAG
\cite{HFAG06} one gets $\acpbskpi\approx +37\%$.
In Fig.~\ref{fig:6} we show
the probability ratio of being $B_s^0$ or $\overline{B}_s^0$.
We measure $\acpbskpi\,=\,+0.39\pm0.15\pm0.08$ with
$2.5\sigma$ significance. While not statistically
significant, this result starts to indicate for the first
time a possible direct CP violation in the $B_s^0$ system. The
size and the sign of the measured asymmetry is in good agreement with
the standard model expectation. 

\section{Conclusions}

We presented here the latest result on the charmless two-body decays
of $B$ hadrons from the CDF experiment. We measure
\acpbdkpi{} which is in agreement with other measurements
and has a comparable uncertainty. More important we observed
three new decays which are \BsKpi, $\Lambda_b^0\rightarrow
pK^-$ and $\Lambda_b^0\rightarrow p\pi^-$. For the decay \BsKpi{}
we measure also the direct CP asymmetry which for the first time
starts to reveal an indication of direct CP violation in the
$B_s$ system.

While we already obtained important results, there is lot of
progress to be expected in this area. First of all, we 
already collected $2.5\ \mathrm{fb^{-1}}$ of data, which is
a substantial increase compared to data used in the presented
results. With this increase in the available statistics we
expect not only a decrease of the statistical uncertainties, but
also the systematic uncertainties as in many cases the dominant
systematic uncertainties come from the limited statistics of
the control data samples. Finally a large sample of
$B_s^0\rightarrow K^+K^-$ decays is interesting for
the lifetime measurement and tagged time dependent
measurements.

\section*{Acknowledgments}

The author would like to thanks the members of the CDF
Collaboration who performed the analysis as well to those who
helped to improve the presentation of this interesting result.


\end{document}